\documentclass[10 pt, conference]{IEEEtran}  % Comment this line out
\usepackage{graphicx} % for pdf, bitmapped graphics files
\usepackage{mathptmx} % assumes new font selection scheme installed
\usepackage{times} % assumes new font selection scheme installed
\usepackage{amsmath} % assumes amsmath package installed
\usepackage{amssymb}  % assumes amsmath package installed
\usepackage{amssymb, bm}
\usepackage[utf8]{inputenc}
\usepackage[T1]{fontenc}
\usepackage{listings}
\usepackage{cite}
\usepackage{multicol}
\usepackage{float}
\usepackage{subfigure}
\usepackage{xcolor}

\IEEEoverridecommandlockouts

\author{\IEEEauthorblockN{Pinyarash Pinyoanuntapong\IEEEauthorrefmark{1},
Tagore Pothuneedi\IEEEauthorrefmark{1},  Ravikumar Balakrishnan\IEEEauthorrefmark{2}, Minwoo Lee\IEEEauthorrefmark{1},  Chen Chen\IEEEauthorrefmark{3}  and
Pu Wang\IEEEauthorrefmark{1}}
\IEEEauthorblockA{University of North Carolina Charlotte\IEEEauthorrefmark{1}, Intel Labs\IEEEauthorrefmark{2} University of Central Florida\IEEEauthorrefmark{3}\\
Email: \{ppinyoan, tpothune,  pu.wang, minwoo.lee@uncc.edu\}\IEEEauthorrefmark{1},
ravikumar.balakrishnan@intel.com\IEEEauthorrefmark{2},
chen.chen@ucf.edu\IEEEauthorrefmark{3}}
 \thanks{This work is funded by Intel/NSF joint grant 2003198 and NSF 2008447}

}

\begin{document}
\title{
Sim-to-Real Transfer in Multi-agent Reinforcement Networking for Federated Edge Computing
}

\maketitle
\pagestyle{empty}

%%%%%%%%%%%%%%%%%%%%%%%%%%%%%%%%%%%%%%%%%%%%%%%%%%%%%%%%%%%%%%%%%%%%%%%%%%%%%%%%
\begin{abstract}

Federated Learning (FL) over wireless multi-hop edge computing networks, i.e., multi-hop FL, is a cost-effective distributed on-device deep learning paradigm. 
%Our recent work on FedEdge used multi-agent reinforcement learning to significantly reduce convergence time and network latency when compared to baseline wireless federated learning systems. FedEdge framework has an unparalleled capability to change into an end-to-end FL system. However, the current challenges with scalable physical testbed hinge on the performance of FedEdge. Adapting to any physical layer simulation has performance drawbacks increasing a gap between simulation and real-world deployments. A realistic physical layer simulator for wireless multi-hop networks can be a cost-efficient and scalable solution. 
This paper presents FedEdge simulator, a high-fidelity Linux-based simulator, which enables fast prototyping, sim-to-real code, and knowledge transfer for multi-hop FL systems. FedEdge simulator is built on top of the hardware-oriented FedEdge experimental framework with a new extension of the realistic physical layer emulator. This emulator exploits trace-based channel modeling and dynamic link scheduling to minimize the reality gap between the simulator and the physical testbed. Our initial experiments demonstrate the high fidelity of the FedEdge simulator and its superior performance on sim-to-real knowledge transfer in reinforcement learning-optimized multi-hop FL.

% an approach to integrate a realistic physical layer wireless multi-hop network simulator (FedEdge simulator) for federated edge computing. FedEdge simulator is a custom-built simulation tool developed for Linux, which provides fast prototyping and sim-to-real knowledge transfer. % that emulates the wireless mesh environment more realistically. 
% Our design of dynamic link scheduling in the FedEdge simulator allows trace-based channel modeling, which is helpful to replay or reproduce a scenario from a real testbed in the simulator and vice versa. We evaluate our approach using three experiments: Training FL, Training MA-RL, and Transfer knowledge from simulation to real testbed to show that our realistic FedEdge simulation bridges the reality gap.

\end{abstract}

%%%%%%%%%%%%%%%%%%%%%%%%%%%%%%%%%%%%%%%%%%%%%%%%%%%%%%%%%%%%%%%%%%%%%%%%%%%%%%%%
\section{INTRODUCTION}
In distributed machine learning, federated learning (FL)\cite{FL_intro} is envisioned as a breakthrough technology, enabling machine learning to work in a distributed manner. In FL, worker nodes compute the model updates locally and send them to the server to update the shared global model. 
This prohibits raw data exchange and reduces potential data privacy risks.

%Multimodel integration allows parallel computation for faster convergence so that robust models can be built under low-latency and low-power requirements.
%This approach significantly reduces the data privacy risk by only sending and receiving the computed local model updates to the server and vice versa instead of sending the data itself. %The model convergence depends on several factors, including the number of workers participating in learning, the data distribution across nodes. More workers lead to parallel computations and multiple iterations, which in general lead to faster convergence. 
%With FL, robust models can be built with low-latency and with low-power requirements.

The edge computing devices interconnected by the wireless multi-hop network constitute the multi-hop wireless edge computing network. Enabling FL over multi-hop wireless edge computing networks (i.e., multi-hop FL) not only can augment AI experiences for urban mobile users, but also can democratize AI and make it accessible in a low-cost manner to everyone, including the large population of people in low-income communities, under-developed regions, and disaster areas. Despite its great  advantages, the convergence of multi-hop FL can be greatly slowed down by the noisy and bandwidth-limited multi-hop wireless links. To address this fundamental challenge, we exploited multi-agent reinforcement learning (MA-RL) for FL optimization, which minimizes the networked induced latency by learning the forwarding paths with the least delay for FL traffic flows \cite{fedair,fededge}. To demonstrate the practical impact of our proposed solution, we developed the FedEdge \cite{fededge}, which is the first experimental framework in the literature for FL over
multi-hop wireless edge computing networks. FedEdge
thus enables fast prototyping, deployment, and evaluation of novel FL algorithms along with machine learning based FL system optimization methods in real-life wireless devices. 

Although FedEdge can provide valuable and broader insights into the practical performance of FL in the field, FedEdge can only run on physical wireless devices. It demandss much higher training time when testing a number of different networking and computing configurations. Scaling the testbed with a more extensive setup makes the computational demand even worse. Simulations provide a cost-efficient and scalable solution to this problem. Running multiple physical layer simulations simultaneously significantly reduces the training time for faster convergence. In addition, high fidelity simulation environment can facilitate the training of the effective reinforcement learning (RL) policies during the exploration stage and enable sim-to-real knowledge transfer \cite{ho2021retinagan}, where the knowledge, e.g., Q networks, learnt from simulations, can be transferred to the real testbed to enhance the efficiency of RL-based approaches.

%FL over multi-hop wireless networks is widely exploited for the advantages in low latency single-hop communications. However, this approach becomes inefficient as the network complexity grows, leading to high operational and maintenance costs. A mesh of wireless interconnected nodes is used as a replacement to alleviate the problems associated with FL over single-hop communications. This provides advantages such as a cost-efficient backbone and low maintenance requirements. The challenges of using FL over traditional multi-hop wireless networks are that the routes are saturated due to the limited bandwidth, causing issues in production environments since a single channel is shared between management and production traffic. Our prior work in FedEdge computing has shown that the use of multi-agent reinforcement learning for edge computing provides valuable insights and results for improving federated learning in wireless multi-hop networks.

% \begin{figure}[t]
% \includegraphics[width=8cm]{fig/sim2real3.pdf}
% \centering
% \caption{Overview of simulation to reality }
% \label{sim2real}
% \end{figure}

\begin{figure}[t]
\includegraphics[width=8cm]{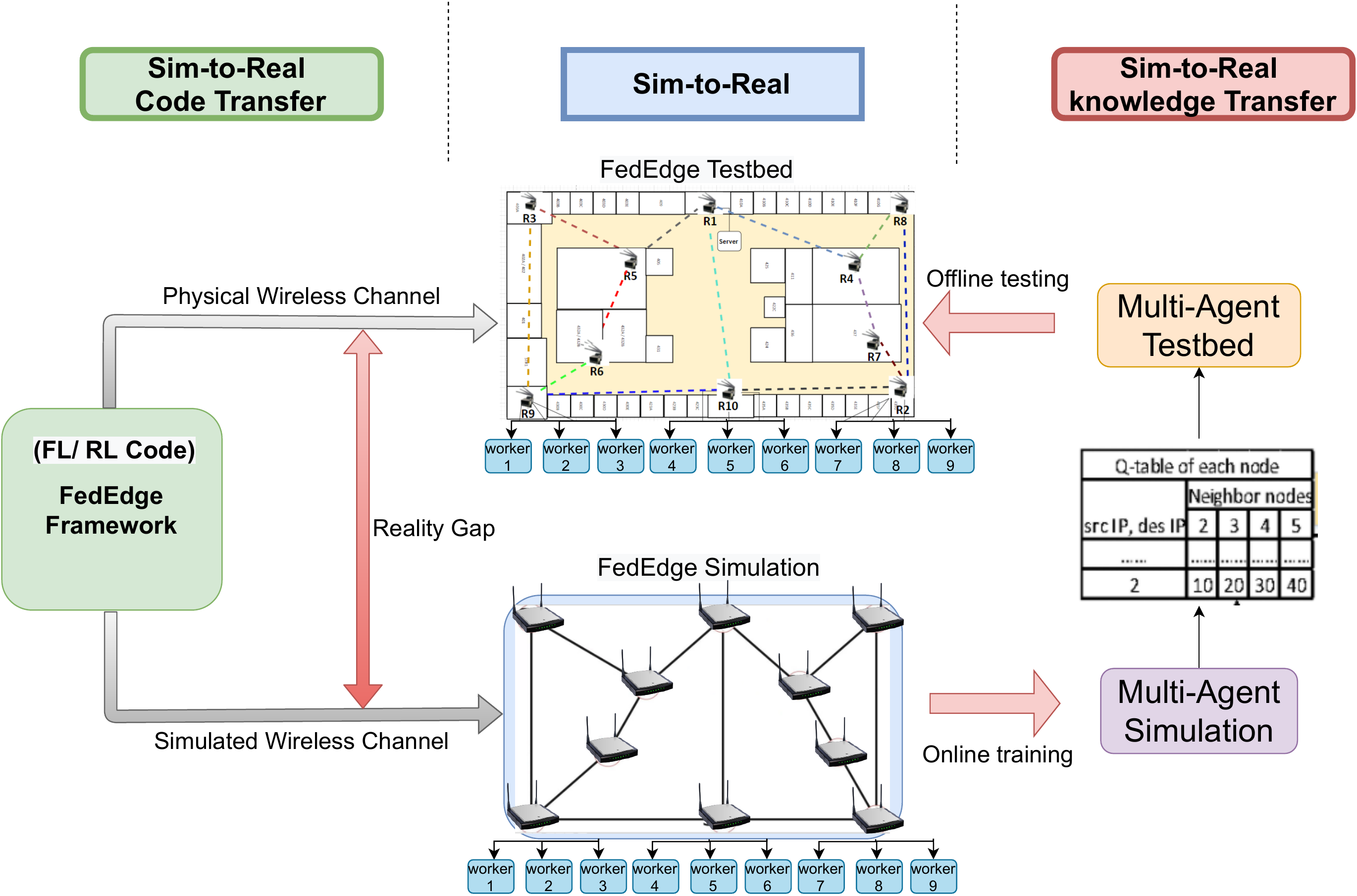}
\centering
\caption{ Overview of Sim-to-Real Transfer }
\label{fig:sim2real}
 	\vspace{-0.6 cm}

\end{figure}

A major impediment is that the only existing network environment simulator to support such requirements is ns3-gym\cite{ns3gym}. ns3-gym is a simulator that integrates the ns-3 and an RL simulation interface, OpenAI gym \cite{gym}. It has a wide range of ML and RL applications to simulate various wired and wireless environments. However, the usage of a real network stack with recent implementations of ns3 using direct execution code (DCE) \cite{dcm} is quite limited due to the lack of significant community effort and limited flexibility/programmability in SDN/open-flow environments. ns-3's implementation of open-flow switch is a class `OpenFlowSwitchNetDevice' and does not have access to the programmable flow tables. 

%So, most of the research community is geared towards using mininet or mininet-wifi, which provides scalable SDN programming flexibility and abstractions to the higher layers. This features ease of code portability to real networks. The challenges with the physical layer implementations in mininet-wifi are that they are static and generalized for wireless environments and fail to support wireless multi-channel based multi-hop backbone networks.

\subsection{Challenges}
Overall, we believe the following challenges in the wireless networking domain limit the progress of federated learning in wireless multi-hop networks.
\begin{itemize}
\item{\emph{Existence of reality gap}}: Even with the overwhelming success of reinforcement learning and federated learning for SD-WAN \cite{fededge}, training the models on simulations and directly executing them on real networks yields poor performance due to the differences between the physical layer simulators and real networks. 

\item{\emph{Lack of reliable training environment}:} 
The performance of machine learning techniques like FL and RL  depends on the training data and a reliable training environment. For many reasons, such as lack of knowledge, time, or resources to build the wireless environment, researchers usually tend to work directly on the testbed setup and spend a large amount of training time to develop models. 

 \item{\emph{Need for wireless multi-hop simulator}:}  The existing wireless simulators mainly support infrastructure-based wireless environments with single-radio single-channel per link which seriously limits the ability to simulate realistic usage settings. They do not account for multi-channel multi-radio based multi-hop network operations which are increasingly becoming realistic. Specifically, using multi-channel multi-radios for wireless mesh backbone networks will tremendously reduce the interference from other nodes and increase the overall performance.
\end{itemize}

Therefore, we build a physical simulator for wireless multi-hop networks called FedEdge simulator to address the current problems of emulating a wireless multi-hop network and integrate it with our FedEdge framework. This approach will give FedEdge the advantage of being the first framework to incorporate a fully functional multi-agent reinforcement wireless network for federated edge computing, which enable to develop and test wireless networks involving federated learning with high fidelity simulations.

\subsection{Contributions}
 
To the best of our knowledge, our implementation is the first attempt in the wireless networking domain to test and validate knowledge transfer from simulation to reality in federated computing with multi-agent reinforcement networking. Our objective in this work is to build a custom realistic physical layer simulator called FedEdge simulator to integrate with our existing FedEdge architecture as shown in Figure~\ref{fed_edge_arch}. Following are our contributions in this paper:
\begin{itemize}
\item \emph{Simulation to real code transfer}:
Using real-time production environments such as OpenFlow software switches and Linux software tools such as mac80211\_hwsim, hostapd, wpa\_supplicant, Linux Namespace Containers (LXC), Traffic Control (TC), and Netlink protocol in the simulator enables rapid development and testing of the code and porting the same codebase to real testbed environments. This reduces the development and debugging times and promotes productivity. 
\item \emph{Simulation to real knowledge transfer}:
Our implementation of the FedEdge simulator using the dynamic link scheduling method reduces the reality gap between the simulations and actual physical networks. Integrated with our existing FedEdge framework, the simulator can help develop accurate models and yield better simulation to real transfer performance. % transfer learning from simulation to real.
% \item{\emph{Scalability}}: Multiple instances of the physical layer simulators can run simultaneously, whereas with a physical testbed, we can run only one experiment at a time.
\item{\emph{Accelerated prototyping for training}}:  Multiple simultaneous runs enable FL and RL to quickly discover hyper-parameters such as
the number of local training round, regularization parameters for model updates, the number of stragglers, model tuning, learning rate, and exploration parameters. FedEdge simulator allows us to evaluate several configurations in parallel thereby speeding up the model selection for the development cycle.

\end{itemize}
\section{Preliminaries}
In this section we briefly provide a background on the fundamental concepts and components considered for our integrated simulation and testbed framework.

\subsection{Accelerating Multi-hop FL via Multi-Agent RL}

A multi-hop FL\cite{fedair,fededge} system consists of a central server that serves as an aggregator and a multi-hop wireless link to edge servers, referred as workers. The architecture of FL across a multi-hop wireless network is shown in Figure \ref{fed_edge_arch}. FL methods are developed to manage distributed neural network training over numerous devices, where each device has its own training data and the goal is to train a common model with an objective to minimize the training loss. 
% Such problem can be modeled as  $ \min _{w} F(w)=\sum_{k=1}^{N} \lambda^{k} F^{k}(w) $
% % $$F^{k}(w)=x^{k} \sim \mathcal{D}^{k}\left[f\left(w^{k} ; x^{k}\right)\right]$$
% where $F(w)$ is the global loss, $F^k(w)$ is the local loss of device $k$, $N$ is the number of devices, $\lambda^k=n^k/n$ and $\sum_{k=1}^N\lambda^k=1$ where $n^k$ is the number of training samples on device \emph{k} and $n=\sum_{k}n^k$ is the total number of training samples in the network. %The local loss $F^k(w)$ is a non-convex function over data distribution $\mathcal{D}^{k}$, which is may vary for devices.

FL methods use a standard optimization technique called local stochastic gradient descent (SGD) to unravel the above optimization problem, which alternates between local SGD iteration and global model averaging for multiple rounds. %, where the worker is the device that participates in collaborative model training.
The worker seeks to decrease its local loss $F^k(w)$ throughout each round by conducting \emph{H} mini-batch SGD iterations. After iterating through \emph{H} local SGD iterations, the worker nodes send their updates to the server, which averages the collected local models and aggregates them into a global model. The updated global model is distributed to the workers, and this process continues again. 
% The communication rounds between the workers and the server are synchronized, i.e., the server waits for all the workers to send their updates before performing the model aggregation. In the worst case, the convergence of the global model can be $
% \mathcal{O}\left(\sqrt{\tau_{\max }} / \sqrt{K t}\right)
% $ where \emph{t} is the training time. $\tau_{\max }$ is the maximum time taken by the slowest worker to send its local model to the server.
Due to the dynamic nature of wireless multi-hop networks,  high and nomadic end-to-end delays become the governing factor in determining the convergence times in multi-hop FL.

% \subsection{Delay Optimal Multi-agent Reinforcement Learning (MA-RL)}

To address such problems, MA-RL was proposed to minimize the wall-clock convergence time to achieve the desired FL accuracy. The problem was formulated as a multi-agent Markov decision Process (MA-MDP) and solved by model-free multi-agent reinforcement learning. In this case, each router is an agent, which observes the network states (e.g., the source IP and destination IP of the incoming FL packet) and learns the optimal networking policy (e.g., the forwarding action at each router) based on the reward signals (i.e., negative per-hop delay) with an objective to maximize the expected total return (i.e., end-to-end delay) from the initial state (i.e., when the FL packet entering the network) to the terminal state (i.e., when the FL packet leaving the network).

 \subsection{FedEdge Experimental Framework}
\begin{figure}[t]
\includegraphics[width=6cm]{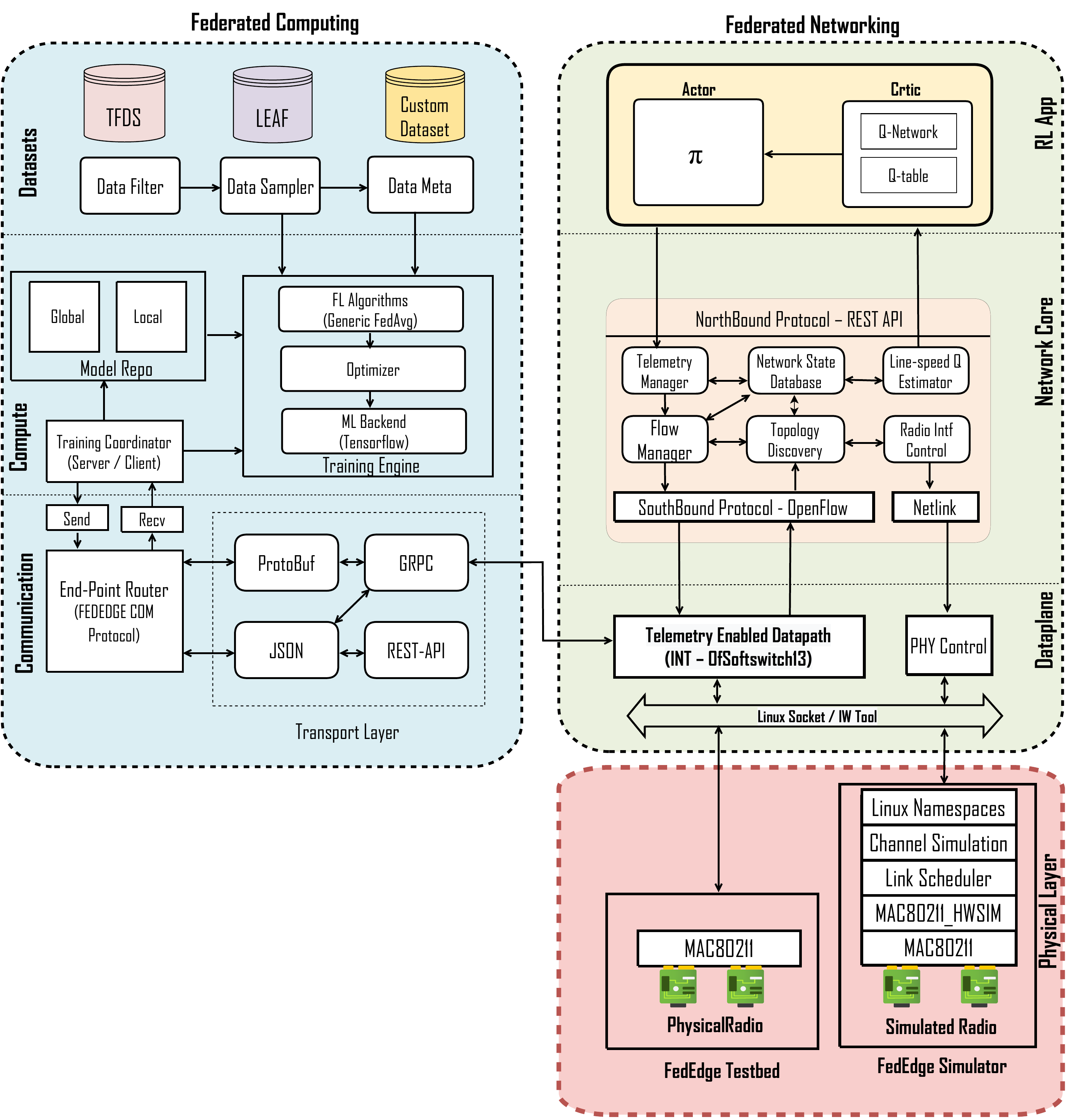}
\centering
\caption{\small FedEdge Framework Architecture with FedEdge~Simulator}
\label{fed_edge_arch}
 	\vspace{-0.4 cm}

\end{figure}

FedEdge (Figure~\ref{fed_edge_arch}) is the experimental prototype framework developed to support wireless multi-hop federated learning. %FedEdge provides modularity with communication functions between the nodes and is easy to integrate into a real testbed or a simulation without any modification to the underlying architecture or the code-base. 
To fast prototype multi-hop FL in the physical wireless devices, we developed FedEdge framework that includes federated computing and federated networking \cite{fededge}. Each component is built in a layered approach where each layer has bidirectional communications to upper or lower layers. In general, federated computing involves customizing and configuring FL functionality, and federated networking is an AI-oriented wireless network operating system that is mainly responsible for fast and reliable wireless network links between the aggregator node and worker nodes. The main goal of this component is to optimize the wireless network through AI-enabled algorithms to perform route optimizations and provide in-band telemetry data. Federated computing contains three layers: (1) Datasets layer that stores the training datasets for FL, (2) Compute layer that provides core functionalities, for instance, to train a model and store the trained model, and finally (3) Communication layer that uses FedEdge COMM protocol to establish and maintain connections with aggregator and worker nodes. Similar to the federated computing component, the federated networking component has three layers that include (1) Dataplane Layer that integrates software switch to allow programmable packet switching and the in-band telemetry,
(2) The Network Core Layer that handles the traditional network functions such as node discovery, maintaining network links, counters and status database, and %as well as managing the traffic flows, etc. 
(3) RL App layer that contains the actor-critic RL agent for learning delay-optimized routes at the edge nodes. The integration of the simulator to FedEdge architecture is simple because of the modularity of FedEdge, which can use the topology built by the FedEdge simulator and train the RL agent on the wireless multi-hop backbone network.

\subsection{FedEdge Physical Testbed}

A software-defined wireless mesh network testbed was deployed in Woodward Hall at the University of North Carolina at Charlotte (UNCC). This testbed consists of 10 Gateworks routers connected with three of WLE900VX wireless interface cards to enable the multi-radio wireless node and run with Ubuntu 20.04 as the operating system. Each mesh router was configured to operate in Mesh Point (MP) mode, with fixed 2.4 and 5 GHz channel, 20 Mhz channel width in 802.11ac operating mode, and 15 dBm transmit power. Then, three wireless interfaces were bridged to a data plane with a programmable packet handling routine (i.e., OpenFlow flow~table). Besides, we deployed 10 of Nvidia Jetson as edge computing nodes.
% Nine nodes serve as local workers to train the federated learning model and a server to aggregate the average model. It can host multiple workers with isolated networks with the configured network namespace on each Jetson node.

% \subsection{Network Simulation}
% Generally, the goal of any wireless network simulation tool is to produce high-performing, realistic, and flexible wireless networks in a system by drastically reducing the deployment costs and time. The end-to-end network simulation includes two stages. The first stage consists of creation of virtual hardware, configuring nodes into access points, base station clients/hosts, and mesh points as well as managing links between the nodes. The second stage includes the medium simulation, medium access simulation and interference. For medium simulations, the simulator will mimic the real channels using signal propagation models. Likewise, while simulating the medium access, the simulator has to consider the co-channel, control the channel interference and back-off timing to mimic the CSMA/CA methodology as in actual wireless environments. Additional features that provide great advantages include software-defined networking (SDN) as well as the ability to act as an environment to train reinforcement learning algorithms. The simulator also allows in generating vast amounts of training data for machine learning applications.

\section{FedEdge Simulator Design and Prototyping}

\subsection{FedEdge simulator overview}
FedEdge simulator is able to mimic electronic hardware wireless radio and its operations to model an environment inside a computer system. It uses virtualized hardware and Linux-based software tools to build and maintain wireless networks. Most of the tools used are Linux in-built tools namely IP namespaces for containerization, mac80211\_hwsim\cite{mac80211hwsim} to create virtual radio interfaces, iw tool to manage the radio interfaces, and batman\_adv\cite{batmanadv} for shortest path routing. The simulator is completely built using the python language and allows the use of open-flow softswitch \cite{openflow_sw} for software-defined networking. In this section, we describe the framework of the FedEdge simulator, the modes of operation, workflow, and channel modeling.

\begin{figure}[t]
\includegraphics[width=6cm]{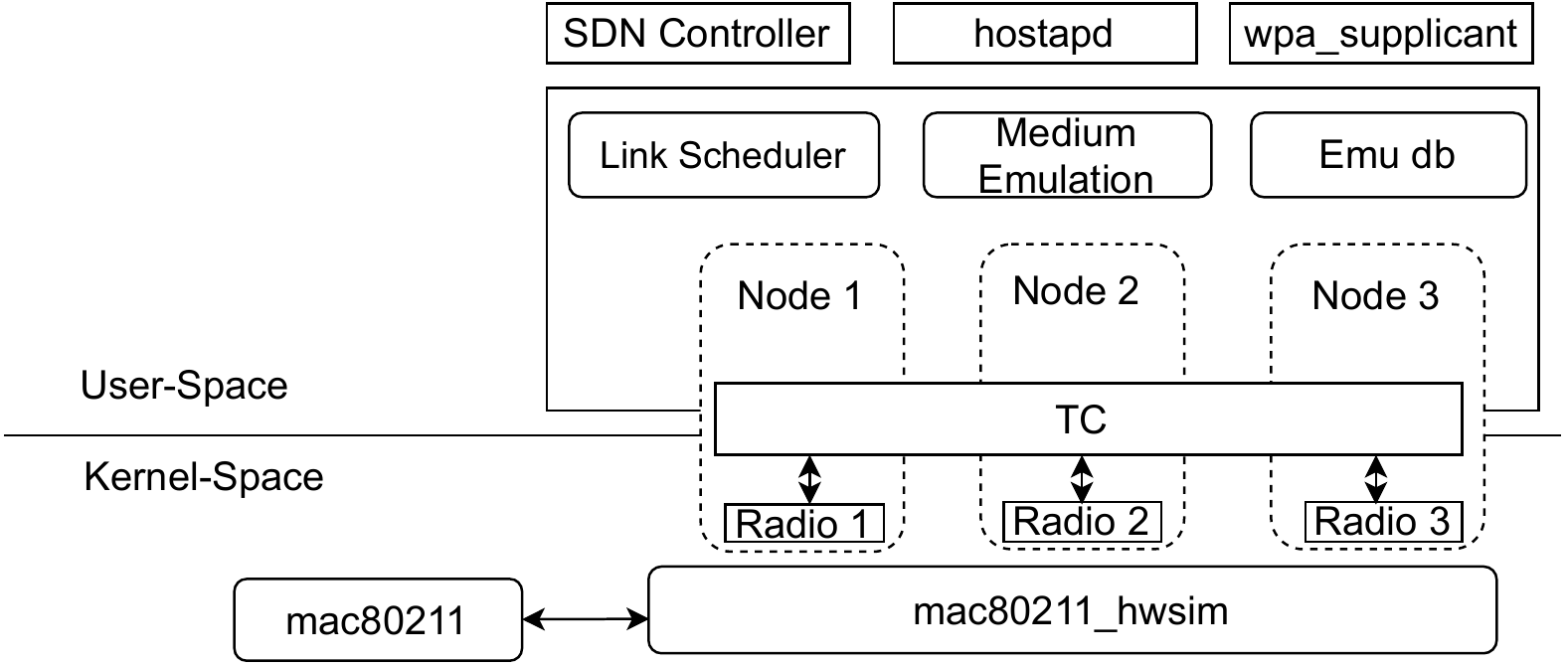}
\centering
\caption{FedEdge Simulator in TClink}
\label{tclink}
 	\vspace{-0.6 cm}

\end{figure}

\begin{figure}[t]
\includegraphics[width=9cm]{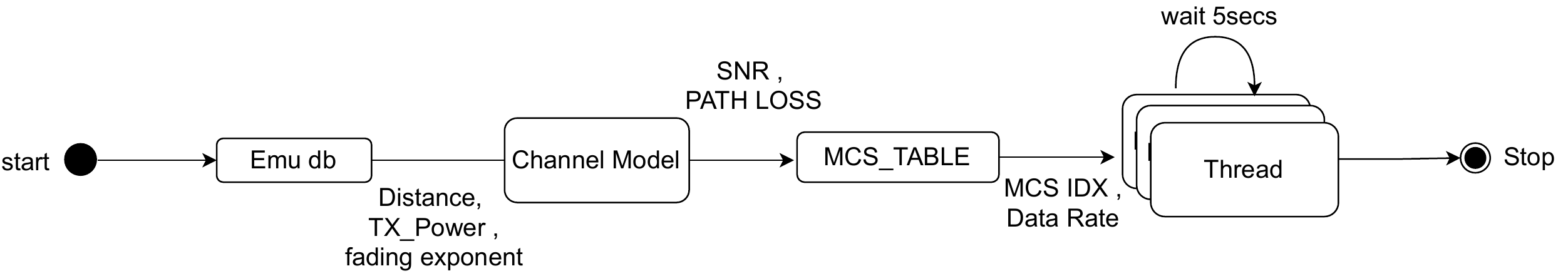}
\centering
\caption{Dynamic Link Scheduling}
\label{dyn_link}
 	\vspace{-0.6 cm}

\end{figure}

As shown in Figure~\ref{tclink}, our simulator is built on top of traffic control (tc) link \cite{tc} to shape the traffic on the egress interface of each node based on the signal-to-noise ratio (SNR) obtained from the channel models and interference models. There is no switching of frames from kernel to user space, ensuring zero copies thereby reducing the overhead. Link scheduler comes into the play in tclink mode to dynamically schedule the links between nodes and runs for every $5$ seconds and interacts with the medium simulator to introduce signal fading and interference between the nodes. 
%The user-space simulator will not receive any messages from the driver, and it does not know active transmissions. So, the trade-off in the tclink mode is to restrict the dynamic interference to static interference where the simulator assumes all the nodes in the same channel are transmitting all the time, thus simulating a worst-case scenario.
%Simulator in TClink mode exclusively uses link scheduler for timely update of link parameters.
Figure~\ref{dyn_link} describes the flow of the link scheduler. Once the simulator starts and builds the necessary wireless medium parameters such as signal fading, interference, and propagation loss, several threads are created for each node in the topology building stage based on the number of the interfaces each node contains.  

\subsection{Simulator Workflow}
\subsubsection{Stages }
 This section describes the general workflow for the FedEdge simulator and its input and output. The input to the simulator is a JSON file that contains context related to the topology. 
 %The simulator works in two stages. Firstly, 
Parsing the input configuration file, the simulator builds the wireless medium parameters and  constructs the topology. The topology is constructed using the mac80211\_hwsim driver, and the links between the nodes are created using the iw tool if the topology is a complete mesh. In addition, the simulator uses wpa\_supplicant and hostapd for client-based connections. If the topology contains a SDN controller setup, the simulator configures the OpenFlow datapath connections on the nodes and connects the nodes to the controller.
%At the second stage, %the simulator works either in Netlink mode or tclink mode. In Netlink mode, the simulator processes the frames from the driver callbacks. In tclink mode,
Next, the simulator uses a link scheduler to update link parameters of each interface. 
%After an interrupt, the topology is cleaned up by deleting the nodes, unloading the drivers, joining threads, cleaning up softswitch processes, etc.

\subsubsection{Channel Modelling}
FedEdge simulator supports multi-hop wireless mesh networks and standard UAV networks. Propagation models available in the simulator are traditional static channel models such as Log Distance, Log-Normal Shadowing, etc. In addition, custom models can be integrated into the simulator that also includes GAN-Based Channel Model \cite{GAN_channel_model} and Trace-Based Channel Model. In stage 1, the channel model name is parsed from the input configuration file. The propagation loss is calculated between the nodes and stored in the simulator database based on the model. In stage 2, the simulator uses the same information to calculate the wireless parameters. %This section discusses the propagation models used in the simulator.
%  \begin{figure}[t]
% \includegraphics[width=8cm]{fig/chan_models_pdf.pdf}
% \centering
% \caption{ Channel Models in FedEdge simulator}
% \end{figure}

% \textbf{\emph{Static Models}} :
% The log distance model\cite{static_channel} is a traditional and static wireless propagation model which is used to predict the propagation loss in various environments. Log Distance model is limited to line-of-sight signal propagation and does not consider obstructions like walls, trees, buildings, etc. The model is based on a simple calculation based on referencing the signal propagation with respect to distances. If path loss from transmitter to distance $d_0$ is $PL_{d_0}$, then the path loss at distance $d$ where $d>d_0$ is given by %the equation below:
% $PL_{d\rightarrow d_0}=PL_0+10nlog{\left(\frac{d}{d_0}\right)}+\chi$
% where $PL_{dO}$ is path loss at distance $d_0$ (dBm), $PL_{d\rightarrow d_0}$ is path loss at distance $d$ (dBm),
% $n$ is path loss exponent depends on the environment, and $\chi$ is normally-distributed random variable. Log-Distance accounts for random shadowing effects by adding a zero-mean Gaussian distribution with standard deviation $\sigma$ expressed in dB.

%The model uses Gaussian distribution random variable to add the shadowing effects to the path loss model.%

\subsubsection{Trace-Based Channel Model} 
Verifying or evaluating a wireless network is the most challenging task given the random, unpredictable nature of the wireless medium. Trace-based channel modeling is an essential technique to replay or reproduce the wireless scenarios from testbed to simulator or vice versa. This approach will provide the required data to model complex wireless systems. Trace-based modeling in the FedEdge simulator works in two modes: replay an existing trace and trace generation for a given input topology. 
% This model works as illustrated in Figure ~\ref{trace_figure}.
% \begin{figure}[t]
% \includegraphics[width=5cm]{fig/trace_replay_gen.pdf}
% \centering
% \caption{ Trace Based Channel Simulation }
% \label{trace_figure}
%  	\vspace{-0.6 cm}
% \end{figure}

\textit{Trace Replay}: 
%\emph{Trace Replay} :
FedEdge simulator can replay the trace for each interface independent of a node. For replaying trace, the simulator checks the topology file for trace file locations and processes the files of each interface. Threads are generated based on the number of replay files and passed to the link scheduler. Based on the signal level on each trace file, the link scheduler runs the traffic control (tc) and sets the bandwidth on each interface of the replay nodes periodically until the trace files are complete. The node's interfaces excluded from the trace replay are manually updated by the simulator and link scheduler based on the input channel model and interference model.

\textit{Trace Generator}:
%\emph{Trace Generate} :
The second mode of the trace model works to generate the trace logs, which can be used to replay on the testbed, visualize or train data for machine learning. When the trace functions are executed, they take the input from the parsed input configuration and learn the replay interfaces. The program works in reverse order compared to replay, where it takes in the interface names and produces the trace files in a csv file format, which contains time, MCS index, RSSI, loss, and traffic rate. The link scheduler captures the data every $5$ seconds until an interrupt occurs. %All the parameters that are recorded to the trace files are calculated by the simulator 
Based on user-selected channel models, interference models, and the topology of the network, the simulator calculates and records all the parameters to the trace files.

\section{Experimental Evaluation}\label{sec:eval}
In this study, we present our experiments in the following scenarios. First, we evaluate the physical layer of FedEdge simulator by comparing the FL experiments alone without MA-RL on both simulator and physical testbed. Next, we compare the reality gap difference by training MA-RL agents in on online fashion in both environments. Last, we study the effect of knowledge transfer of the MA-RL by evaluating the pre-trained model's performance in the physical testbed. % performs on the physical testbed. % with the pre-trained Q-table from the simulator.

% In this study, we present our experiments in two different methods which will demonstrate the usage of the emulations. Then, we briefly discuss about the choice of the topology and feasibility of the experiments. With the physical layer computational gap coming from the testbed some experiments seem far more effective compared to real network. To balance the external factor corresponding to computational and hardware overheads we choose to demonstrate our results in two ways which will likely show the clear picture.

\subsection{Experiment Setup}

\textbf{Network Environment Setup:} To evaluate the gap between the simulated network and the real testbed environment, we identically set up the topology with the same wireless conditions with 10 routers, 9 workers, and a server as shown in Figure~\ref{fig:sim2real}. %Consequently, they have an identical setup for testbed and simulation. 
We set up our FedEdge framework to work on top of the physical layer of the simulator by using shortest path routing and MA-RL to route packets. A replay simulation runs by inputting the RSSI signal logs from the testbed backbone nodes in WiNSLAB at UNCC. The simulator uses the link scheduler and updates the bandwidth on the link interfaces based on the signal seen at the point of the replay time. The link scheduler converts the signal to traffic rate based on MCS index table of 802.11ac 20MHz channels specified at \cite{mcs}.

\textbf{Model and Dataset:}   The CNN model has two convolution layers, with 32 and 64 filters respectively. Each convolutional layer is attached with a 2x2 max pooling layer. The convolutions are followed by a fully connected layer with 128 units with ReLU activation. The final output layer consists of a fully connected layer with sofmax activation. The learning rate of 0.1 is used for local SGD in all workers and the  model size is $5.8$ Mbytes. 
We test the model with the well-known FEMNIST benchmark dataset, the extended federated version of MNIST \cite{mnist} from the LEAF\cite{leaf}, which consists of 62 classes. %, benchmark that is well-known for federated learning. 
%FEMNIST consists of 62 classes.

\begin{figure}[ht]
	\centering
	\subfigure[Iteration loss convergence]{
		\includegraphics[width=0.45\linewidth]{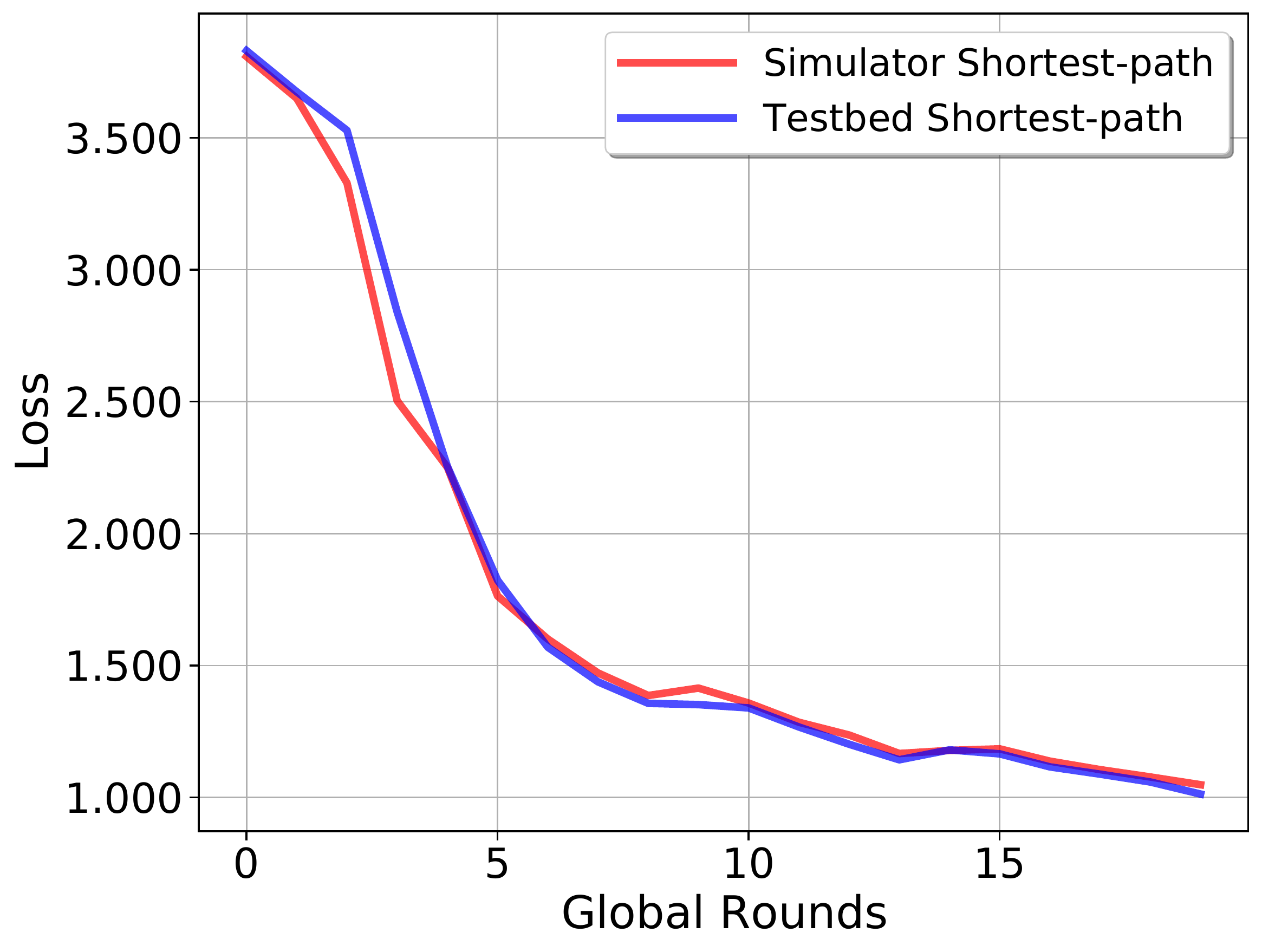}
		\label{fig:short-conv}
	}
	\subfigure[Wall-clock time loss convergence]{
		\includegraphics[width=0.45\linewidth]{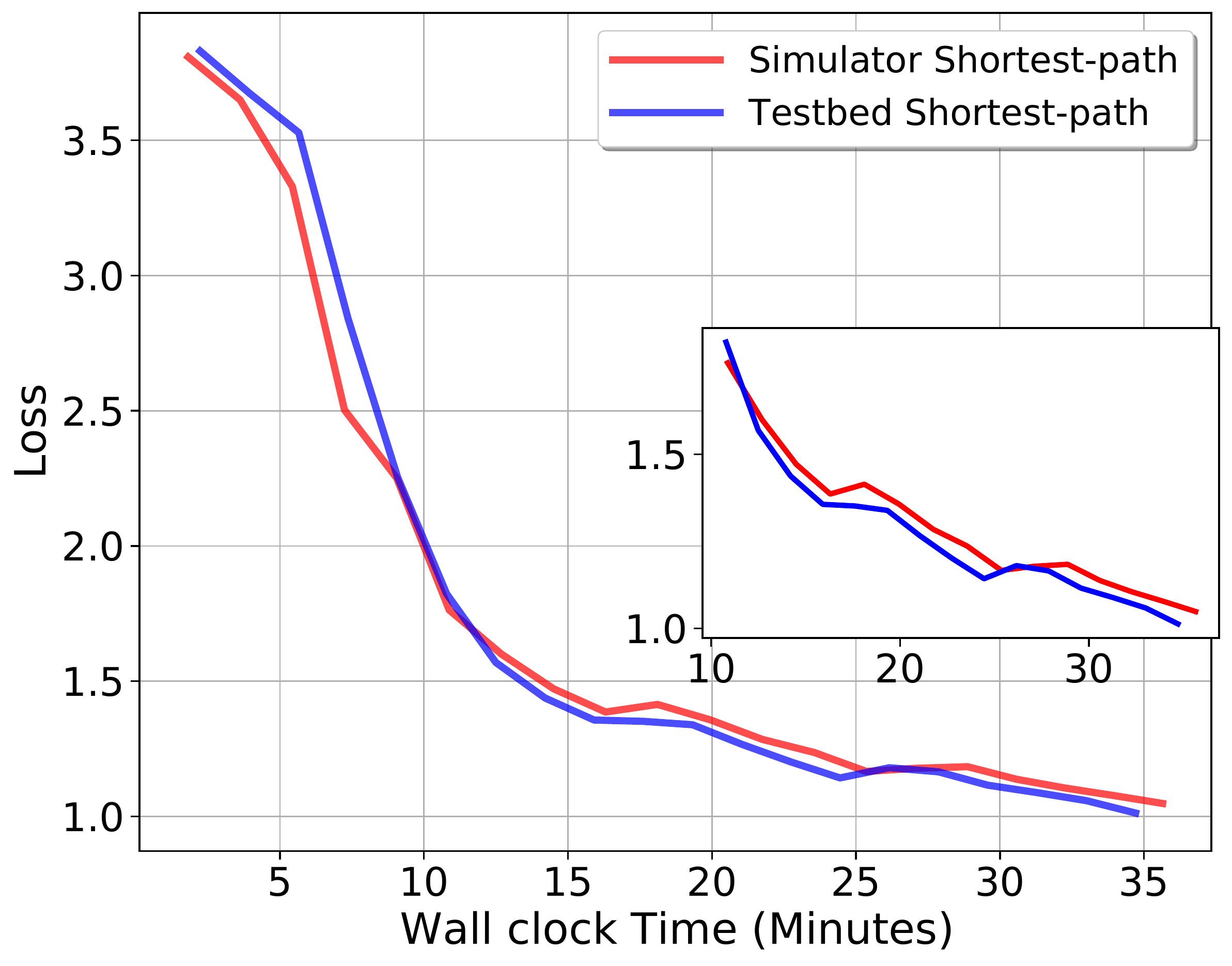}
		\label{fig:short-time}
	}

	\caption{Loss convergence of FL after 20 epochs of shortest-path routing in FedEdge simulator (red) and on testbed (blue)} 
	\label{fig:short}

\end{figure}

\subsection{Main Results}
\textbf{Simulator Fidelity Evaluation:}
Figure~\ref{fig:short} examines the closeness of the realistic physical layer in the proposed simulation to the physical testbed with the experimental results of shortest-path routing in both environments. Both FL experiments lead to the same iteration convergence performance at which they achieve the same loss after running the same number of epochs (Figure~\ref{fig:short-conv}) and slightly different wall-clock time (Figure~\ref{fig:short-time}). Therefore, we verify that both environments are almost identical. 
    
\begin{figure}[ht]
 	\vspace{-0.3cm}

	\centering
	\subfigure[Iteration loss convergence]{
		\includegraphics[width=0.45\linewidth]{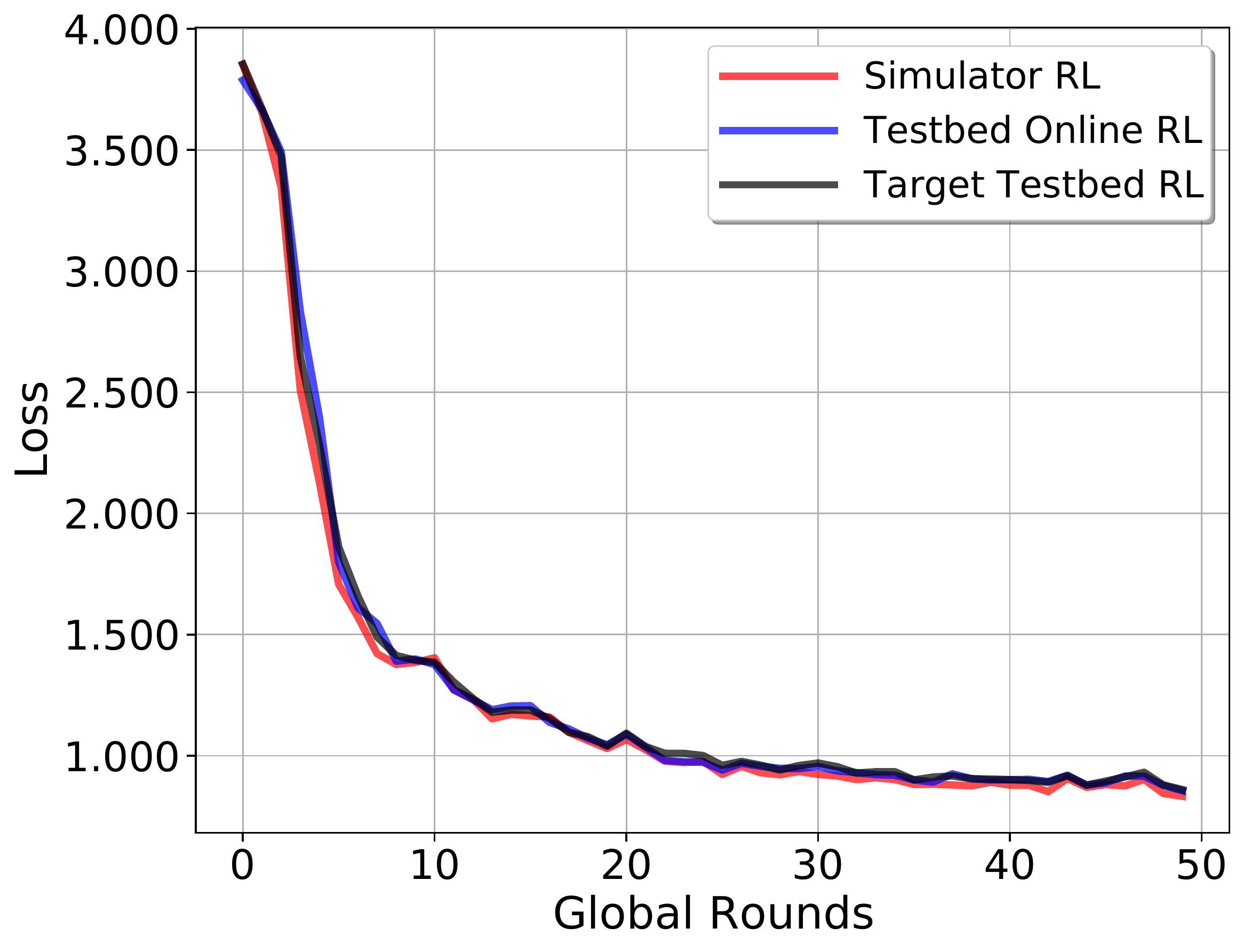}
	}
	\subfigure[Wall-clock time loss convergence]{
		\includegraphics[width=0.45\linewidth]{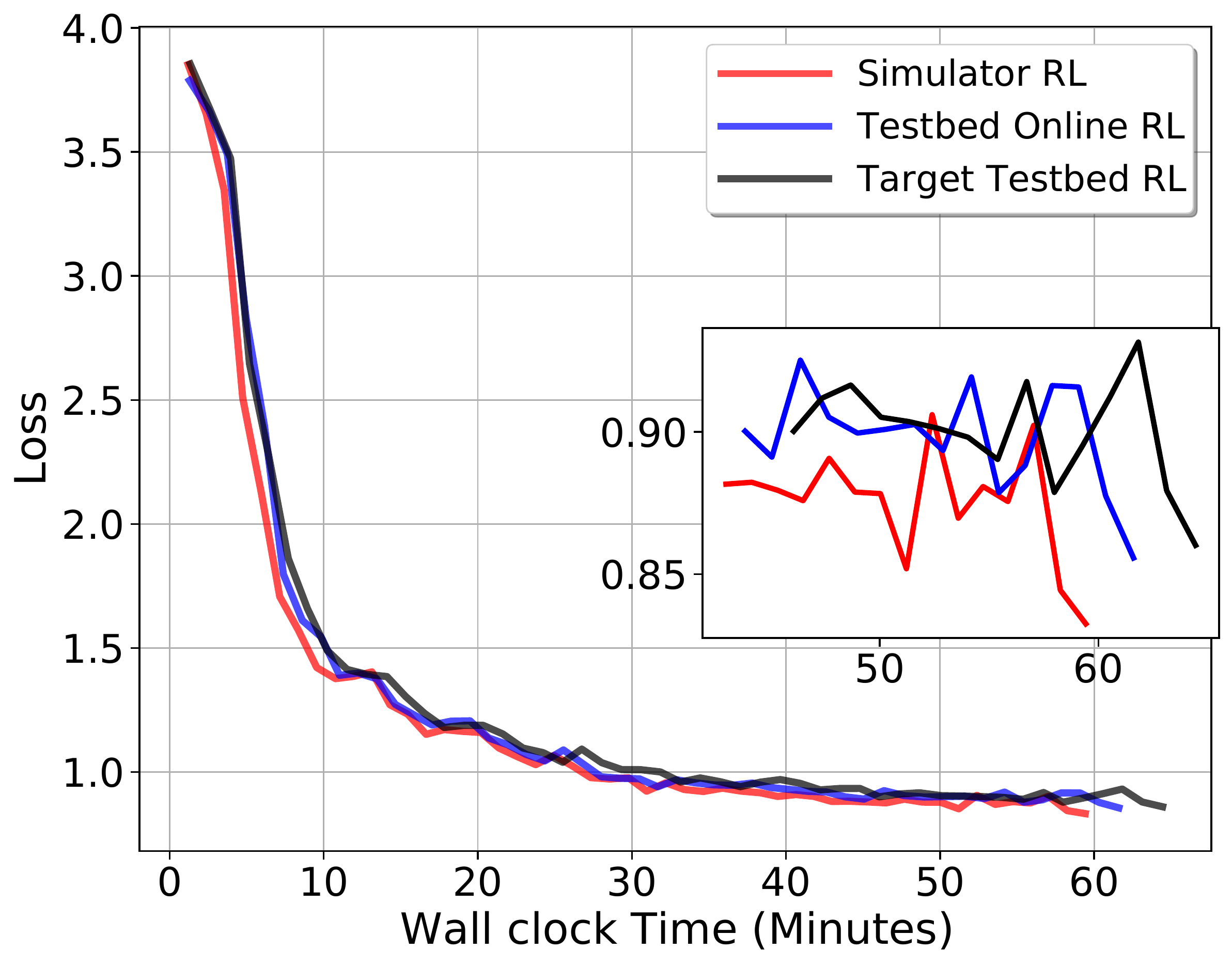}
	}

	\caption{Loss convergence comparison results after 50 epochs of On-policy softmax in FedEdge simulator (red), On-policy softmax on Testbed Online learning (blue), On-policy softmax on Testbed Target testing (black)} 
	\label{fig:transfer_loss}
 	\vspace{-0.6cm}

\end{figure}

%After we verified that both of environments are almost identical from the above experiment with shortest path, we ran 
In the FL experiment with RL-based routing, where each agent is trained in an online fashion in simulator and physical testbed, the iteration loss convergence in terms of global rounds are identical as shown in Figure~\ref{fig:transfer_loss}. However, an online RL routing in the simulator achieves slightly better wall-clock time than in the testbed after 50 epochs, which is only 2 minutes difference. 
Given the identical protocol stacks for both, the results confirm that the FedEdge simulator narrows the reality gap effectively. 
%Our FedEdge implementations uses the same underlying protocol stacks in both simulator and physical testbed. 
%This illustrates that the proposed FedEdge simulator can closely bridge the gap between simulated physical wireless environment and real dynamic wireless environment. 

\begin{figure}[ht]
	\centering
	\subfigure[Time per round]{
		\includegraphics[width=0.45\linewidth]{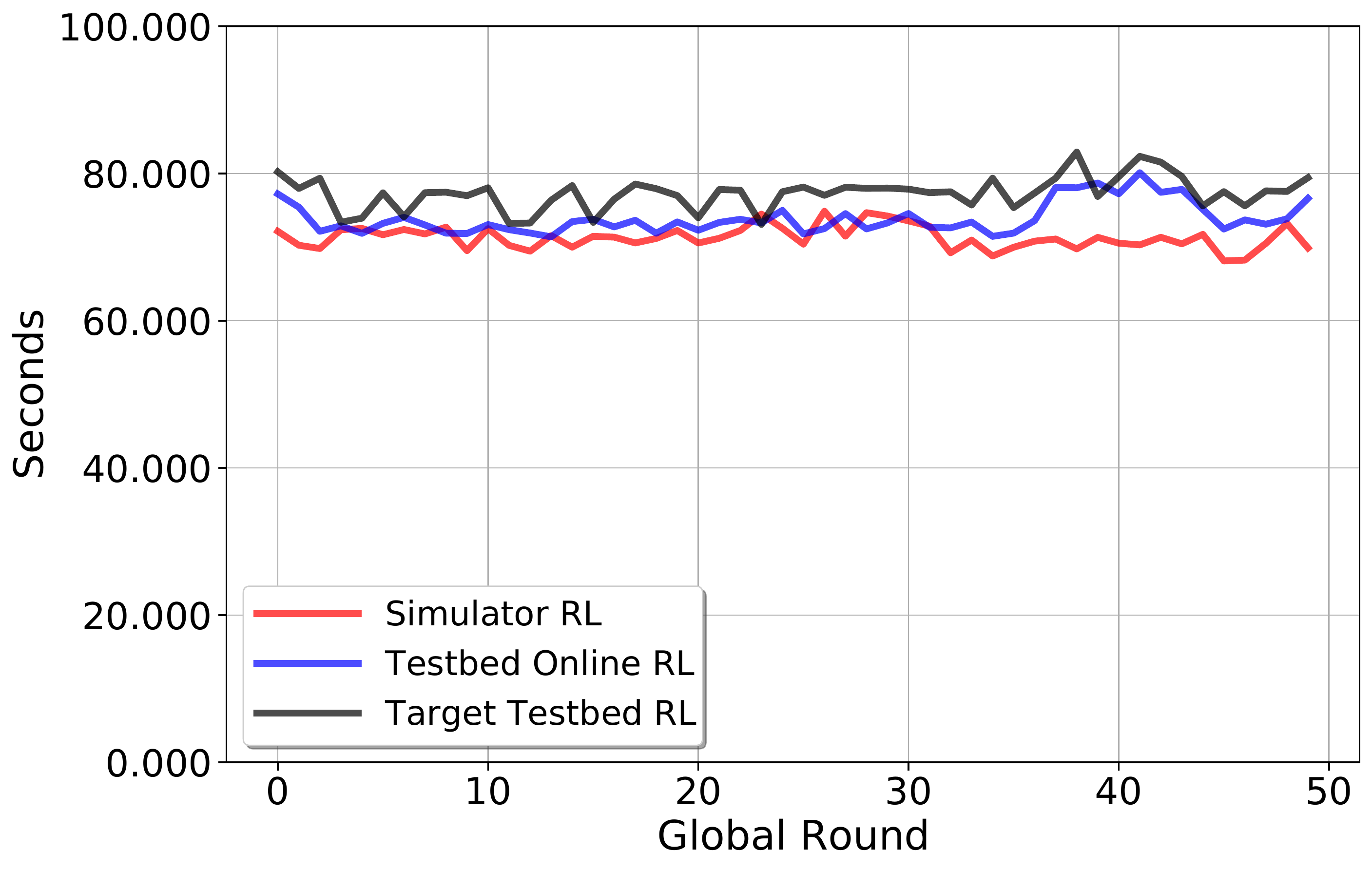}
		\label{fig:transfer_time-round}
	}
	\subfigure[Average time per round]{
		\includegraphics[width=0.45\linewidth]{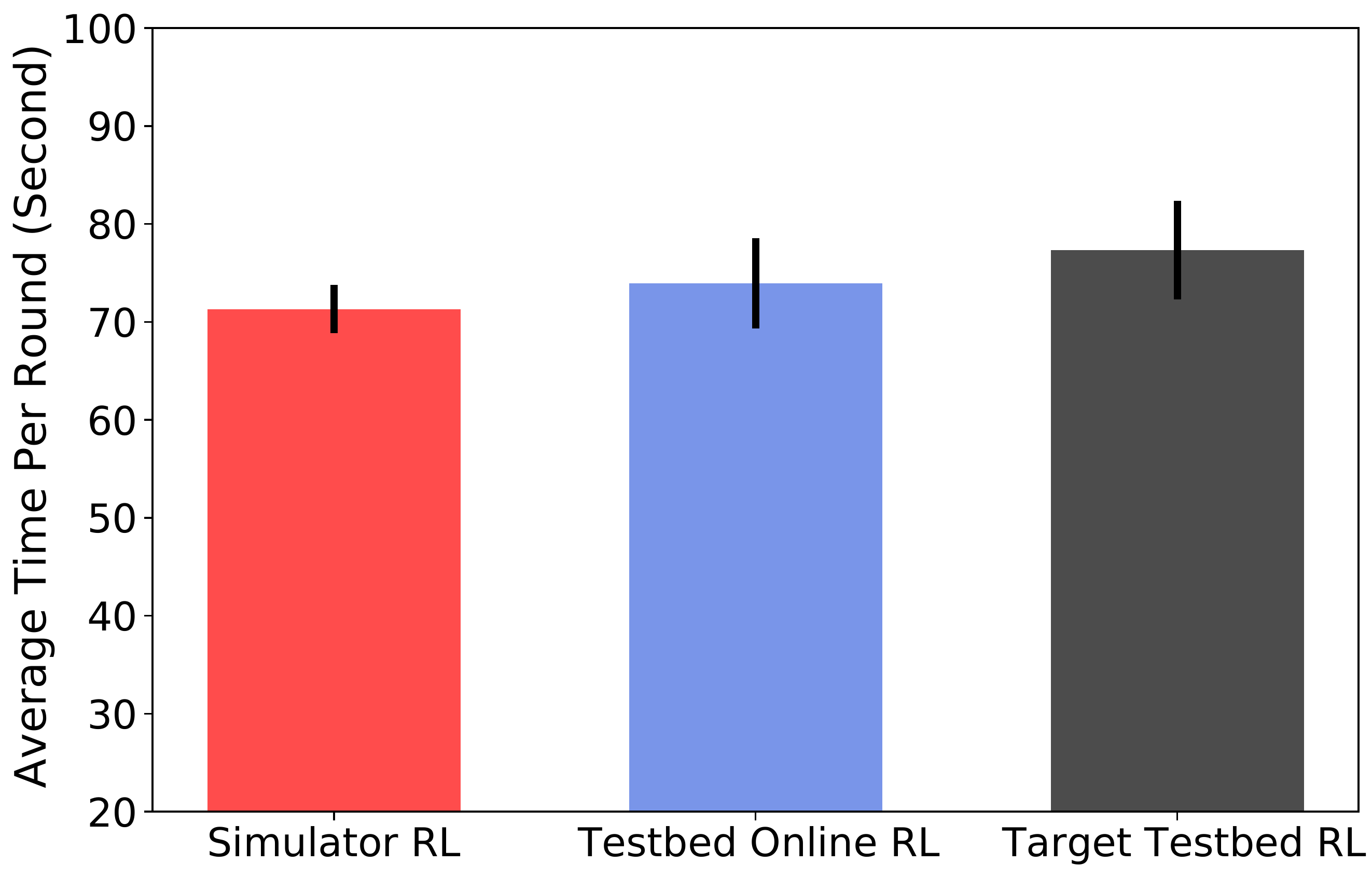}
		\label{fig:transfer_time-bar}
	}

	\caption{Time and average time per round of On-policy softmax in FedEdge simulator (red), On-policy softmax on Testbed Online learning (blue), On-policy softmax on Testbed Target testing (black) across 50 rounds of FL communications} 
	\label{fig:transfer_time}

\end{figure}
 	\vspace{-0.2cm}

\textbf{MA-RL Sim-to-real Knowledge Transfer:}
Now, we evaluate the initial sim-to-real transfer performance of the online MA-RL routing given the verified close reality gap.
%We performed the parallel training at which we train an online training of RL in simulation and testbed environment. 
We pre-train MA-RL agents in a simulated environment and transfer the learned knowledge (Q-table of each agent) to the testbed for the target testing, where we freeze the Q-table and allow an agent to only exploit the pre-trained softmax policy. In Figure~\ref{fig:transfer_loss}, we observe that online training both on the simulator and testbed achieve a better performance in terms of wall-clock time compared to the target testing in the testbed. This is because the frozen target testing lost adaptability to re-learn the dynamic FL traffic patterns in the wireless environment. However, the gap between the target testing in the testbed and both online trainings is marginal. Figure~\ref{fig:transfer_time-round} shows that training time per epoch for both testbed experiments follow the same trend, which are around 72-80 seconds per epoch and slightly slower than the online simulator. Figure~\ref{fig:transfer_time-bar} shows that the average time and variance per epoch of all the approaches across 50 rounds of FL communications. 
% The average time and variance of online training in simulator is slightly lower than the online training and offline testing in testbed because there are still some channel interference from the real network environment.

% \begin{figure*}[t]
% 	\centering
% 	\subfigure[Iteration loss convergence]{
% 		\includegraphics[width=0.30\linewidth]{result/transfer_loss.pdf}
% 	}
% 	\subfigure[Wall-clock time loss convergence]{
% 		\includegraphics[width=0.30\linewidth]{result/transfer_loss_time.pdf}
% 	}
% 	\subfigure[Average time per round]{
% 		\includegraphics[width=0.30\linewidth]{result/avg_time_bar.pdf}
% 	}
% 	\caption{comparison results after 50 epochs of On-policy softmax in FedEdge simulator (red), On-policy softmax on Testbed Online learning (blue), On-policy softmax on Testbed Offline testing } 
% 	\label{fig:rl_sm}
%  	\vspace{-0.4cm}

% \end{figure*}

% \begin{figure}[t]
% \includegraphics[width=8cm]{result/transfer_loss.pdf}
% \centering
% \caption{ Trace Based Channel Simulation }
% \label{fig:transfer_round}

% \end{figure}

\section{CONCLUSIONS}
MA-RL knowledge transfer in FL over wireless multi-hop network is challenging due to the reality gap between the code base and entire stack from the physical layer to the application layer. In this paper, we presented a fully functional approach for transfer learning from simulation to reality in wireless federated edge computing networks. Moreover, we implemented a real physical testbed to validate the proposed FedEdge simulator for effective sim-to-real transfer. Our results confirm the high fidelity of the proposed FedEdge simulator that significantly minimizes the gap from real testbed. %The offline training strategy was able to achieve the similar result comparable to online training in the testbed.

\addtolength{\textheight}{-12cm}   % This command serves to balance the column lengths

%%%%%%%%%%%%%%%%%%%%%%%%%%%%%%%%%%%%%%%%%%%%%%%%%%%%%%%%%%%%%%%%%%%%%%%%%%%%%%%%

%%%%%%%%%%%%%%%%%%%%%%%%%%%%%%%%%%%%%%%%%%%%%%%%%%%%%%%%%%%%%%%%%%%%%%%%%%%%%%%%

%%%%%%%%%%%%%%%%%%%%%%%%%%%%%%%%%%%%%%%%%%%%%%%%%%%%%%%%%%%%%%%%%%%%%%%%%%%%%%%%

\bibliographystyle{IEEEtran}
{\footnotesize
\bibliography{bib/bib-champ, bib/bib-prabhu}}

% \bibliography{bib/bib-lee-marl,bib/FL_pu,bib/bib-Chen,bib/UW_Pu,bib/SAS_Pu,IEEEabrv,bib/bib-prabhu}

\end{document}